\documentstyle[12pt,aasms4]{article}
\slugcomment{To appear in Astrophysical Journal Letters}

\lefthead{Montes et al.}
\righthead{Radio Detection of SN 1986E}

\begin{document}

\title{Radio Detection of SN 1986E in NGC 4302}
%\title{NOT A PREPRINT -- DO NOT DISTRIBUTE \\
%Radio Detection of SN 1986E in NGC 4302}
\author{Marcos J. Montes}
\affil{Remote Sensing Division, Naval Research Laboratory, Code 7214,
Washington, DC 20375-5320; mmontes@moon.nrl.navy.mil}

\author{Schuyler D. Van Dyk\altaffilmark{1}}
\affil{Dept. of Physics \& Astronomy, UCLA, Los Angeles, CA 90095; vandyk@jean.astro.ucla.edu}

\author{Kurt W. Weiler}
\affil{Remote Sensing Division, Naval Research Laboratory, Code 7214,
Washington, DC 20375-5320; kweiler@SNe.nrl.navy.mil}

\author{Richard A. Sramek}
\affil{P.O. Box 0, National Radio Astronomy Observatory, Socorro, NM
87801; dsramek@nrao.edu}

\and

\author{Nino Panagia\altaffilmark{2}}
\affil{Space Telescope Science Institute, 3700 San Martin Drive, Baltimore, MD 
21218; panagia@stsci.edu}

\altaffiltext{1}{Visiting scientist.}
\altaffiltext{2}{Affiliated with the Astrophysics Division, Space
Science 
Department of ESA.}

\begin{abstract}
Radio observations of SN 1986E have shown a clear detection of
emission at 6 cm wavelength about 8 months after optical discovery.
Combined with a number of new upper limits and a study of the possible
models, it appears that SN 1986E was probably a fairly normal Type IIL
supernova, somewhat similar to SN 1980K, with radio emission at
roughly expected levels.  This detection continues the correlation
between radio detection and late time optical emission.
\end{abstract}

\keywords{supernovae, SN 1986E}

\section{Introduction}

Discovered by G. Candeo at m$_{pg}$ = 14.5 on 1986 April 13 in NGC
4302 (Rosino 1986), with a precise optical position of RA(B1950) =
$12^h 19^m 09.^s 05$, Decl(B1950) = $+14^{\circ} 54' 33.''3$ given by
King (1986), SN 1986E attracted little initial attention from
observers. Recently, interest in the SN has revived with discovery of
late time optical emission (Cappellaro, Danziger, \& Turatto 1995), a
relatively rare phenomenon for supernovae (SNe), leading to a
comparison with other late time optical emitting SNe 1957D, 1970G,
1979C, and 1980K (Cappellaro et al. 1995; see also Ryder et al. 1993
for SN 1978K, and Rupen et al. 1987 and Leibundgut et al. 1991 for SN 1986J).  Due to these
optical similarities and, since all are radio supernovae (RSNe),
Cappellaro et al. (1995) predicted possible radio emission for SN 1986E
with flux density as high as 0.5 - 5 mJy at 20 cm in 1995, $\sim$9
years after optical discovery.  This prediction was followed by a
search for radio emission in 1995 by Eck et al. (1996), who were unable
to detect the SN with very low limits at both 6 cm (S$_{\rm 6cm}$ $<$
0.038 mJy; 3$\sigma$)$ $ and 20 cm (S$_{\rm 20cm}$ $<$ 0.169 mJy;
3$\sigma$).  Due to this apparent lack of radio emission, Eck et al.
(1996) concluded that SN 1986E is surprising in being apparently radio
faint and suggested that it is the first old SN that has been seen in
the optical but not the radio.

This work by Cappellaro et al. (1995) and by Eck et al. (1996) has
prompted us to reexamine observations taken with the Very Large
Array\footnote{The VLA is operated by the NRAO of the AUI under a
cooperative agreement with the NSF.} (VLA) of
SN 1986E at times closer to the explosion date, one of which had showed
a possible, weak detection.  Applying additional processing, we were
able to obtain a clear detection of SN 1986E at 6 cm on 1986 December
05 and to establish a number of new upper limits at both 6 and 20 cm.
These new data constrain rather well the range of possible models for
the radio emission from SN 1986E, so that a comparison with other Type
IIL SNe such as SN 1979C and SN 1980K can be made.  From these results
we conclude that the Cappellaro et al. (1995) predicted flux density
for SN 1986E is far too high and, while SN 1986E was probably a
reasonably typical Type IIL SN in its radio emission, that Eck et al.
(1996) failed to detect it due to limited VLA sensitivity and the
relatively large age of the SN at the time of their search.

\section{Observations}

A number of observations were taken of SN 1986E at 6 cm wavelength
(4.860 GHz) and 20 cm (1.425 GHz) with the VLA over an interval
extending from 17 days to 991 days after the optical discovery.  The
observations were calibrated with assumed flux densities for the
primary calibrator 3C286 of 14.45 Jy at 1.425 GHz and 7.42 Jy at 4.860
GHz.  Following the normal procedure for the VLA, 3C286 was used to
obtain flux densities for the compact secondary, possibly variable,
calibrator 1252+119, which was then used as both amplitude and phase
calibrator for observations of the SN 1986E field.  (More details of
VLA calibration procedures for SN observations can be found in Weiler
et al. 1986 and, particularly for 1252+119, in Weiler et al. 1991.)
The derived flux densities for 1252+119 used for amplitude calibration
are listed in Table 1 and the position of 1252+119 used for phase
calibration was RA(B1950) = $12^h 52^m 07.^s 717$, Decl(B1950) =
$+11^{\circ} 57' 20.''86$.  The results of the observations of SN 1986E
are listed, along with the two limits from Eck et al. (1996), in Table
1.

As is apparent from Table 1, one of the observations of SN 1986E
resulted in a weak, but significant detection of 0.304 $\pm$  0.034 mJy
at 6 cm on 1986 December 05.  The radio position of the emission of
RA(B1950) = $12^h~ 19^m~ 09.^s 09$, Decl(B1950) = $+14^{\circ}~ 54'~
32.''5$ agrees with the optical (King 1986) position to within 1''.
This positional coincidence, along with the lack of detection of any
similar level of emission both before and after the single detection
date, indicates that the emission almost certainly is from SN 1986E.  A
6 cm contour map of the area around the position of SN 1986E is shown
in Figure 1 with the emission from the SN indicated.

Two relatively strong background sources, labeled $\beta$ and $\gamma$
by Eck et al. (1996), are clearly seen in Figure 1.  The Eck et al.
(1996) source $\delta$, while outside of the field shown in Figure 1,
was also detected.

The 6 cm flux densities for these three field sources, taken as an
average of the results from 1986 August 11 (B-array) and 1986 December
05 (C-array) are listed in Table 2.  An observation from 1987 August
28 was not used for Table 2 because the high resolution of the A-array
on that date partially resolved the sources.  The positions for the
three field sources (Source $\beta$: RA(B1950) = $12^h~ 19^m~ 09.^s83$, Decl(B1950) =
$+14^{\circ}~ 54'~ 51.''3$; Source $\gamma$: RA(B1950) = $12^h~ 19^m~ 14.^s06$, Decl(B1950) =
$+14^{\circ}~ 53'~ 35.''0$; Source $\delta$: RA(B1950) = $12^h~ 19^m~ 01.^s31$, Decl(B1950) =
$+14^{\circ}~ 53'~ 50.''7$) are in good agreement with the results of
Eck et al. (1996).

\section{Parameterized Model}

Weiler et al. (1986) discuss the common properties of radio SNe (RSNe),
including nonthermal synchrotron emission with high brightness temperature, 
turn-on delay at longer wavelengths, power-law decline after maximum 
with index 
$\beta$, and spectral index $\alpha$ asymptotically decreasing to 
an optically thin value.  Weiler et al. (1986) have also shown
that the ``mini-shell'' model 
of Chevalier (1982a,b) adequately describes known Type IIL RSNe.  In 
this model, the 
relativistic electrons and enhanced magnetic fields necessary for synchrotron
emission are generated by the SN shock 
interacting with a relatively high-density ionized circumstellar envelope.
This dense cocoon is presumed to have been
established by a high mass-loss rate ($\dot M$~ $>$~ 10$^{-6}$ $M_\odot$~
yr$^{-1}$), low velocity ($w_{\rm wind}$~ $\sim$ 10 ~km ~s$^{-1}$) wind from a
red supergiant (RSG) SN progenitor which was ionized and 
heated by the initial SN UV/X-ray flash.
The rapid rise in radio flux density results from the shock 
overtaking progressively more of the wind matter, leaving less of it along 
the line-of-sight to absorb the emission from the shock 
region.

Following Weiler et al. (1986), we adopt the parameterized model: 
\begin{equation} 
S {\rm (mJy)} = K_1 {\left({\nu} \over {\rm 5~GHz}\right)^{\alpha}}
{\left({t - t_0} \over {\rm 1~day}\right)^{\beta}}
e^{-{\tau}} 
\end{equation}
where
\begin{equation} 
\tau = K_2 {\left({\nu} \over {\rm 5~GHz}\right)^{-2.1}} {\left({t -
t_0} 
\over {\rm 1~day}\right)^{\delta}} 
\end{equation}
\noindent with $K_1$ and $K_2$ corresponding, formally, to the flux 
density ($K_1$) and uniform absorption ($K_2$) at 5~GHz one day after 
the explosion date $t_0$.  The term 
$e^{-{\tau}}$ describes the attenuation of a local medium that 
uniformly covers the emitting source (``uniform external
absorption'').  This absorbing medium is assumed to be purely 
thermal, ionized hydrogen with frequency dependence $\nu^{-2.1}$.  The 
parameter $\delta$ (not to be confused with Source
$\delta$ from Eck et al. 1996) describes the time dependence of
the optical depth for this local, uniform medium.  For an
undecelerated SN shock, $\delta$ = -3 is appropriate (Chevalier, 1982a).

This parameterization has been found generally applicable to other
Type IIL SNe, such as SN 1979C (Weiler et al. 1991) and SN 1980K
(Weiler et al. 1992), with values of $\delta$ close to the
undecelerated value ($\delta_{\rm SN1979C}$ = -3.12, $\delta_{\rm SN1980K}$ 
= -2.74, see Table 3).

\section{SN 1986E Model Parameter Values/Limits}

With only the one radio detection available in Table 1, it is
certainly not possible to determine the 6 parameters ($K_1$, $K_2$, 
$\alpha$, $\beta$, $\delta$, $t_0$)
in Eqs. 1 and 2 through a fitting procedure.  However, if we
assume no shock deceleration ($\delta$ = -3), if we assign
a ``typical'' spectral index for a Type IIL RSN of $\alpha = -0.65$
($\alpha_{\rm SN1979C}$ = -0.73, $\alpha_{\rm SN1980K}$ =
-0.58, see Table 3), and if we take $t_0$ = -30 days before
optical maximum (optical discovery date in the case of SN 1986E), as we did
for SN 1980K (Weiler et al. 1992), only $K_1$, $K_2$, and $\beta$ remain
undetermined.  

This is still too many parameters to determine unambiguously from only
one detection, but through the use of the 3$\sigma$ upper limits
available at 6 cm on 1986 August 11 (4 months before the
detection) and on 1987 August 28 (8 months after the detection),
we can place limits on their possible values.  These are listed in 
Table 3 and the resulting model radio light curves are shown along with
the available data in Figure 2.

Accepting these limits for $K_1$ and $\beta$, the two parameters which
predict the late time radio behavior, implies that SN 1986E would have
had a flux density $<$ 0.060 mJy at 20 cm on 1995 July 14 and 
$<$ 0.026 mJy at 6 cm on 1995 December 04, the epochs of the Eck et 
al. (1996) observations.  These are both less than their reported 3$\sigma$
upper limits (see Table 1) and consistent with their
non-detection of the SN in the radio at those times and frequencies.

\section{SN 1986E Estimated Properties}

Using these ``best estimate'' parameters for SN 1986E and taking a
distance to NGC 4302 of 16.8 Mpc (Tully 1988) yields a 6 cm peak spectral luminosity of $L_{\rm 6cm~ peak}$ $\simeq$ 1.1 $\times$ 10$^{26}$ erg s$^{-1}$
Hz$^{-1}$.  Although more than an order of magnitude less radio
luminous than SN 1979C at peak ($L_{\rm 6cm~ peak} = 2.6 \times
10^{27}$ erg s$^{-1}$ Hz$^{-1}$, see Table 3), SN 1986E is comparable
in 6 cm peak luminosity to SN 1980K ($L_{\rm 6cm~ peak}$ = 1.0 $\times$
10$^{26}$ erg s$^{-1}$ Hz$^{-1}$, see Table 3).

Further, using the parameters from Table 3 and the formulation of
Weiler et al. (1986), Eq. 16, we can obtain an estimate of the ratio
of the presupernova mass loss rate ($\dot M$) to presupernova stellar wind
velocity ($w_{\rm wind}$) of $\frac {\dot M}{w_{\rm wind}}~ >$
4.7 $\times$~ 10$^{-6}$ ~$M_\odot$ yr$^{-1}$ (km~ s$^{-1}$)$^{-1}$ or
$\dot M~ >$ 4.7 $\times$~ 10$^{-5}$ ~$M_\odot$ yr$^{-1}$ ($\dot M_{\rm
SN1979C}$ = 1.9 $\times$~ 10$^{-4}$ ~$M_\odot$ yr$^{-1}$, $\dot M_{\rm
SN1980K}$ = 2.0 $\times$~ 10$^{-5}$ ~$M_\odot$ yr$^{-1}$, see Table 3),
for the commonly assumed parameters of $w_{\rm wind} =$ 10 km s$^{-1}$,
T = 20,000 K, and $v_{\rm shock} =$ 13,000 km s$^{-1}$.

A comparison of the model and physical parameters derived for SN
1986E, SN 1979C, and SN 1980K is shown in Table 3.  As is clear from
examination of Table 3, SN 1986E was likely a fairly typical Type IIL
SN with properties somewhat similar to those of SN 1980K.

\section{Conclusions}

Although our radio data sample is extremely limited with only one
detection at one frequency, by including new upper limits it is
possible to show, with reasonable assumptions and parameter estimates,
that SN 1986E was probably a fairly ``normal'' Type IIL SN in its radio
behavior.  The best parameter estimates imply that the late time radio
upper limits determined by Eck et al. (1996) are not in conflict with
the estimated properties of SN 1986E, and their conclusion that SN
1986E represents the first example of a late time optically detectable
SN which has not been seen in the radio is not supported.  SN 1986E
could, in fact, have been rather similar to SN 1980K.  Thus, the strong
correlation between the presence of radio emission and late time
optical emission is maintained.

\acknowledgments

KWW \& MJM wish to thank the Office of Naval Research (ONR) for the 6.1
funding supporting this research.

\clearpage

\begin{deluxetable}{ccccccc}
\footnotesize
\tablecaption{Radio Observations of SN 1986E. \label{tbl-1}}
\tablewidth{0pt}
\tablehead{
\colhead{Observation} & \colhead{Days from} 
& \colhead{VLA} & \colhead{Flux Density}
& \colhead{Error} & \colhead{Frequency} 
& \colhead{Calibrator} \\
\colhead{Date} & \colhead{Reference Date} & \colhead{Config.} &
\colhead{(limits are $3\sigma$)}
& \colhead{} & \colhead{} & \colhead{1252+119} \\
\colhead{} & \colhead{} & \colhead{} & \colhead{(mJy)} & \colhead{(mJy)} & \colhead{(GHz)} 
& \colhead{(Jy)}
}
\startdata
86/04/13 & $\equiv$ 0.00 \nl
86/04/30 &  17     & A  &$<$0.330  &0.110  &4.860  &0.632 \nl
86/06/26 &  74     &A/B &$<$0.255  &0.085  &4.860  &0.632 \nl
86/08/11 & 120     & B  &$<$0.203  &0.068  &4.860  &0.632 \nl
86/08/11 & 120     & B  &$<$0.540  &0.180  &1.425  &0.938 \nl
86/12/05 & 236     & C  &   0.304  &0.034  &4.860  &0.635 \nl
86/12/05 & 236     & C  &$<$0.630  &0.210  &1.425  &0.954 \nl
87/08/28 & 502     & A  &$<$0.180  &0.060  &4.860  &0.606 \nl
88/12/29 & 991     & A  &$<$0.150  &0.050  &4.860  &0.594 \nl
95/07/14\tablenotemark{a} &3379     & A  &$<$0.169  &0.056  &1.425  &0.754 \nl
95/12/04\tablenotemark{a} &3522     & B  &$<$0.038  &0.013  &4.860  &0.748 \nl
\enddata
\tablenotetext{a}{From Eck et al. (1996).}
\end{deluxetable}

\clearpage

\begin{deluxetable}{ccc}
\footnotesize
\tablecaption{Flux Density at 6 cm for Field Sources.\label{tbl-2}}
\tablewidth{0pt}
\tablehead{
\colhead{Source} & \colhead{Flux Density\tablenotemark{a}} &
\colhead{Flux Density} \\
\colhead{Name} & \colhead{} & \colhead{Error} \\
\colhead{} & \colhead{(mJy)} & \colhead{(mJy)}
}
\startdata
$\beta$  & 1.70 & 0.11 \nl
$\gamma$ & 1.78 & 0.11 \nl
$\delta$ & 0.27 & 0.07 \nl
\enddata
\tablenotetext{a}{Flux densities are an average from
measurements on 1986 August 11 and 1986 December 05.}
\end{deluxetable}

\clearpage

\begin{deluxetable}{cccc}
\footnotesize
\tablecaption{Comparison of SN 1986E with SNe 1979C and 1980K. \label{tbl-4}}
\tablewidth{0pt}
\tablehead{
\colhead{Parameter} & \colhead{SN 1986E} 
& \colhead{SN 1979C\tablenotemark{a}} & \colhead{SN 1980K\tablenotemark{a}}
}
\startdata
K$_1$       & $>$120               & 1800  & 120 \nl
$\alpha$\tablenotemark{b}    & $\equiv$-0.65        & -0.73 & -0.58 \nl
$\beta$     & $<$-1.03             & -0.81 & -0.73 \nl
K$_2$       & $>$3.9 ~$\times$~ 10$^6$ & 1.1 ~$\times$~ 10$^8$ & 1.9
~$\times$~ 10$^5$ \nl
$\delta$\tablenotemark{b}    & $\equiv$ -3.00        & -3.12 & -2.74 \nl
t$_0$\tablenotemark{b}~(days) & $\equiv$ -30          & $\equiv$ -15 & $\equiv$ -30 \nl
L$_{\rm 6cm~ peak}$~(erg~s$^{-1}$~Hz$^{-1}$) & $\simeq$1.1 ~$\times$~ 
10$^{26}$ & 2.6 ~$\times$~ 10$^{27}$ & 1.0 ~$\times$~ 10$^{26}$ \nl
$\dot M$\tablenotemark{b} (M$_\odot$~yr$^{-1}$) & $>$4.7 $\times$ 10$^{-5}$ & 1.9 ~$\times$~
10$^{-4}$ & 2.0 ~$\times$~ 10$^{-5}$ \nl
\enddata
\tablenotetext{a}{New data and model fit have yielded slightly
different values than Weiler et al. (1991, 1992).}
\tablenotetext{b}{See text for assumptions.}
\end{deluxetable}

\clearpage

\begin{figure}
\plotfiddle{86efield9.eps}{300pt}{0}{75}{75}{-300}{-30}
\caption{VLA 6 cm detection of SN 1986E from 1986 December 05.
The Eck et al. (1996) sources $\beta$ and $\gamma$ are readily apparent.
Their source $\delta$ (see Table 2) is outside
of this field.  Contour levels are -1 (dashed), 1.5, 2, 2.5, 3, 5, 10,
20, 30, and 50 $\times$ 10$^{-1}$ mJy beam$^{-1}$.}
\label{fig1}
\end{figure}

\clearpage

\begin{figure}
\plotfiddle{86efit9.eps}{300pt}{0}{90}{90}{-250}{-120}
\caption{The single detection (shown with error bar) at 6 cm
and upper limits (arrows) at 6 cm (squares) and 20 cm (stars) are shown
with model radio light curves (6 cm curve solid; 20 cm curve
dashed).}
\label{fig2}
\end{figure}

\clearpage


\begin{thebibliography}{}
\bibitem[Cappellaro, Danziger, \& Turatto 1995]{kin95} Cappellaro, E.,
Danziger, I. J., \& Turatto, M. 1995, \mnras, 277, 106
\bibitem[Chevalier 1982a]{che82a}Chevalier, R.A. 1982a, \apj, 259, 302
\bibitem[Chevalier 1982b]{che82b}Chevalier, R.A. 1982b, \apj, 259, L85
\bibitem[Eck et al. 1996]{eck96} Eck, C. R., Cowan, J. J., Boffi,
F. R., \& Branch, D. 1996, \apjl, 472, L25
\bibitem[King 1986]{kin86} King, D. L. 1986, \iaucirc ~4206
\bibitem[Leibundgut et al. 1991]{leib91} Leibundgut, B., Kirshner,
R. P., Pinto, P. A., Rupen, M. P., Smith, R. C., Gunn, J. E.,
Schneider, D. P. 1991, \apj, 372, L531
\bibitem[Rosino 1986]{ros86} Rosino, L. 1986, \iaucirc ~4202
\bibitem[Rupen et al. 1987]{rup87} Rupen, M. P., van Gorkom, J. H.,
Knapp, G. R., Gunn, J. E., \& Schneider, D. P. 1987, \aj, 94, 61
\bibitem[Ryder et al. 1993]{ryd93} Ryder, S., Staveley-Smith, L.,
Dopita, M., Petre, R., Colbert, E., Malin, D., \& Schlegel, E. 1993, \apj, 416, 167
\bibitem[Tully 1988]{tul88} Tully, R.B. 1988, Nearby Galaxies
Catalogue, Cambridge Univ. Press
\bibitem[Weiler et al. 1986]{wei86} Weiler, K.W., Sramek, R.A.,
Panagia, N., van der Hulst, J.M., \& Salvati, M. 1986, \apj, 301, 790
\bibitem[Weiler et al. 1991]{wei91}Weiler, K.W., Van Dyk, S.D.,
Panagia, N., Sramek, R.A., \& Discenna, J.L. 1991, \apj, 380, 161
\bibitem[Weiler et al. 1992]{wei92}Weiler, K.W., Van Dyk, S.D.,
Panagia, N. \& Sramek, R.A. 1992, \apj, 398, 248
\end{thebibliography}
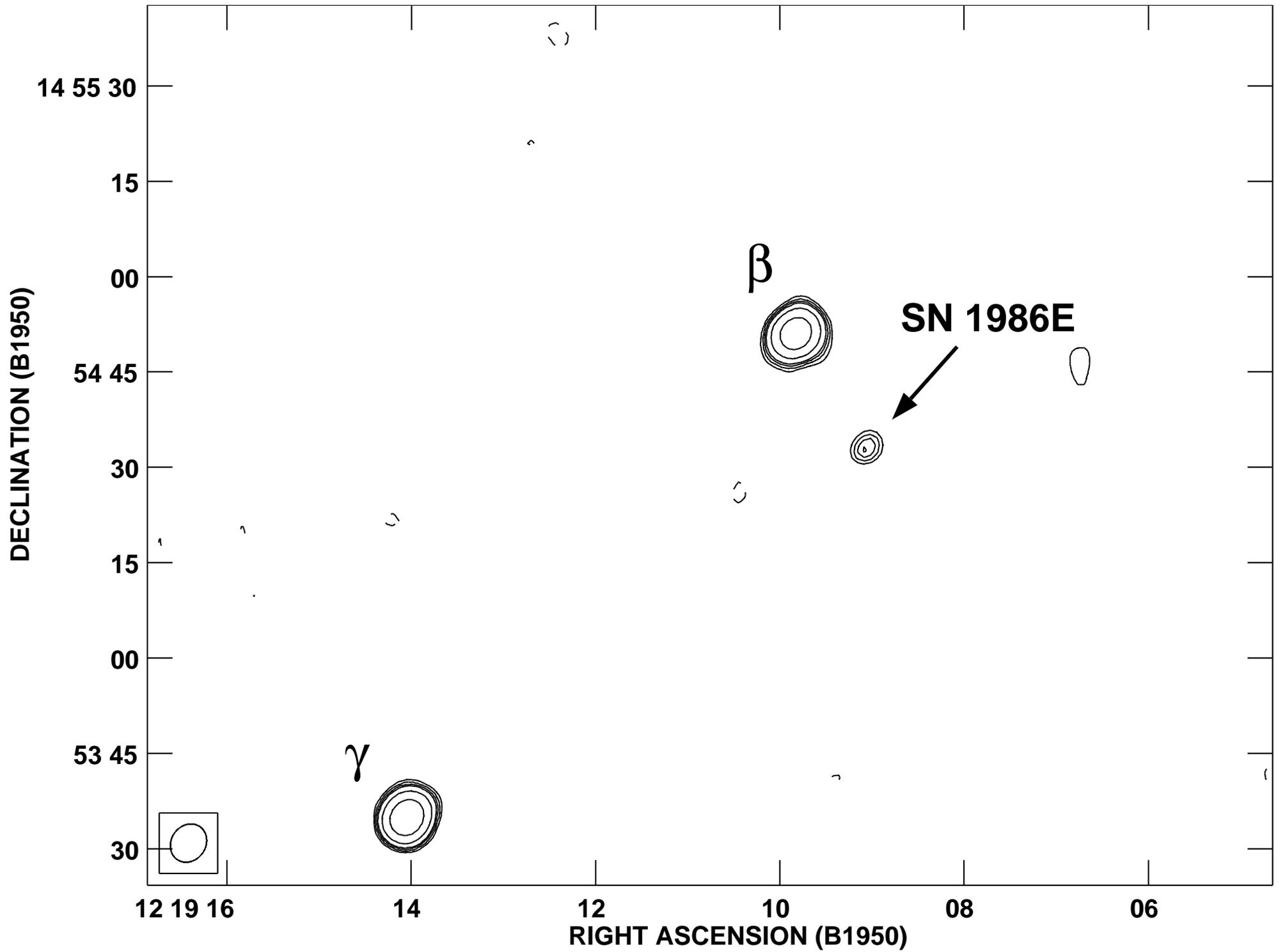
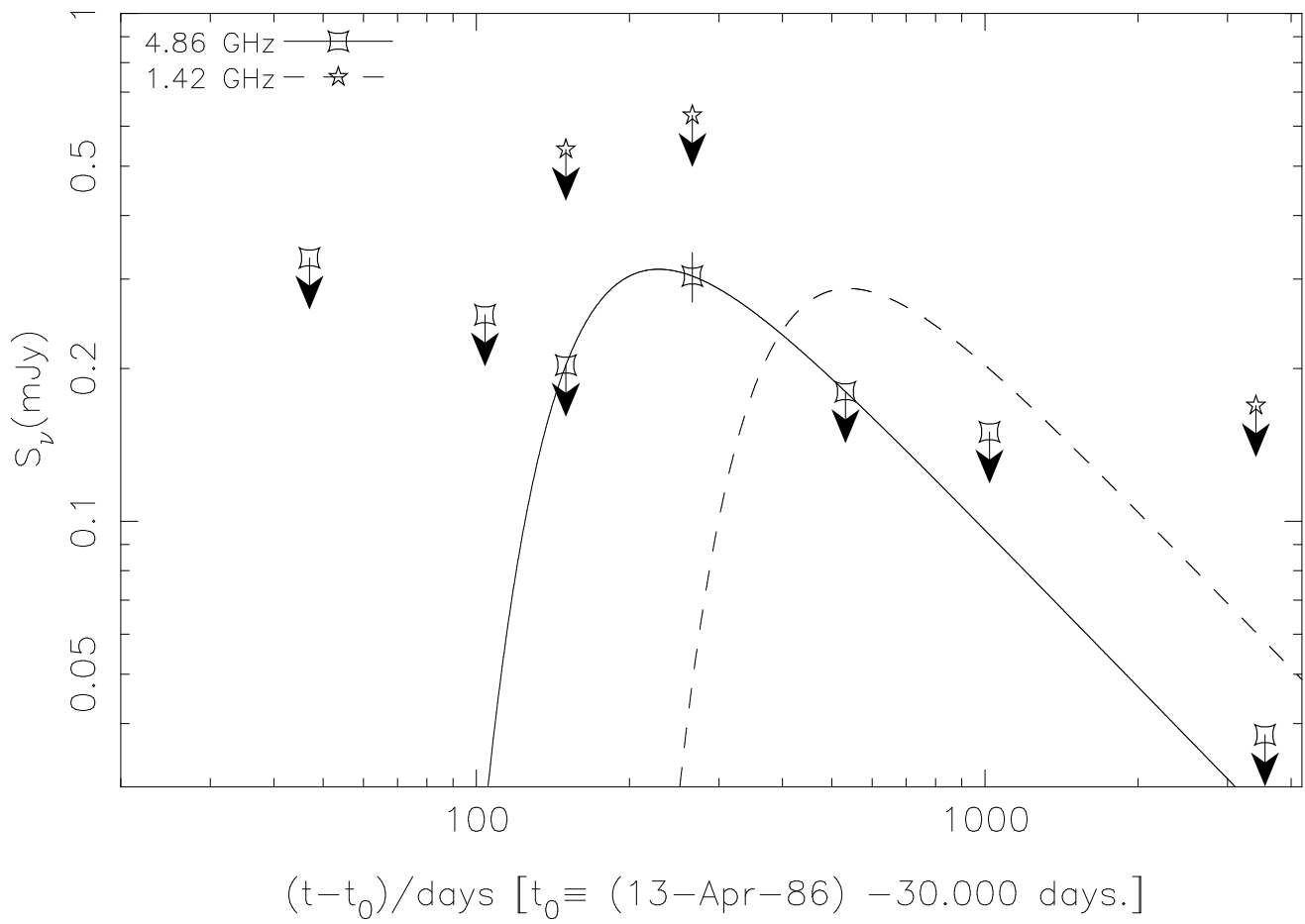
\end{document}